# KitRobot: A multi-platform graphical programming IDE to program mini-robotic agents

Nadeem Akhtar

PhD - IRISA – University of South Brittany - FRANCE

Assistant Professor

Department of Computer Science and IT, The Islamia University of Bahawalpur

Baghdad-ul-Jadeed campus, Pakistan。 Tel: +92-331-2116491, E-mail: nadeem.akhtar@iub.edu.pk

Anique Akhtar

Electrical Engineering, Graduate School of Science and Engineering

Koc University, Acarlar 26/10, Sariyer, Istanbul, Turkey。 E-email: akhtaranique@gmail.com

*The research is financed by Department of Computer Science and IT, The Islamia University of Bahawalpur, Pakistan*

**Abstract:**
The analysis, design and development of a graphical programming IDE for mini-robotic agents allows novice users to program robotic agents by a graphical drag and drop interface, without knowing the syntax and semantics of the intermediate programming language. Our work started with the definition of the syntax and semantics of the intermediate programming language. The major work is the definition of grammar for this language. The use of a graphical drag and drop interface for programming mini-robots offers a user-friendly interface to novice users. The user can program graphically by drag and drop program parts without having expertise of the intermediate programming language. The IDE is highly flexible as it uses xml technology to store program objects (i.e. loops, conditions) and robot objects (i.e. sensors, actuators). Use of xml technology allows making major changes and updating the interface without modifying the underlying design and programming.

## 1. Introduction

KitRobot is a platform independent graphical interface designed and developed to program a mini-robot. The purpose of this system is to program graphically without learning the syntax and semantics of the underlying robot low-level language. It creates reliable, reusable, and maintainable program code rapidly, and abstracts away the technical complexity of the underlying low-level robot procedural language from the user. This in turn allows the novice users without having the knowledge and expertise of the low-level robot procedural language to develop efficient programs. A multi-agent robotic system is distributed. A distributed system and the interactions between components within it exhibit a high-level of complexity and lead to difficulty in the assessment of what system behavior is possible in different scenarios. Constraining such a system requires us to fully understand the system behavior, and to place controls on sets of activities the system can perform. An agent is considered as a computer system situated in some environment, capable of autonomous actions in this environment in order to meet its design objectives [Wooldridge and Jennings, 1995]. Multiple agents are necessary to solve a problem, especially when the problem involves distributed data, knowledge, or control. A multi-agent system is a collection of several interacting agents in which each agent has incomplete information or capabilities for solving the problem [Jennings et al., 1998].

These are complex systems and their specifications involve many levels of abstractions. They have concurrency, and often have dynamic environments. To catch up with the complexity problems in them and get significant results with formal analysis, we must cope with complexity at every stage of development: from the specification phase to the analysis, design and verification phase. The formal specification and verification of these systems along with step-wise refinement from abstract to concrete concepts play a major role in system correctness. Safety and liveness properties have to be enforced during each development phase of requirement specifications, verification specifications, architecture specifications, and implementation.

Our multi-agent robotic system [Akhtar, 2010] is composed of robotic transporting agents. The global mission is to transport material from one storehouse to another. They move in their environment which is static i.e. the topology of the system does not evolve at run time.

There are three types of agents
1) *Carrier agent:* agent that transports stock material from one storehouse to another, it can be loaded or unloaded i.e. full or empty, and can move both forward and backward. Each road section is marked by a sign number and the carrier agent can read this number. Each carrier agent has a sensor to detect a collision with another carrier.

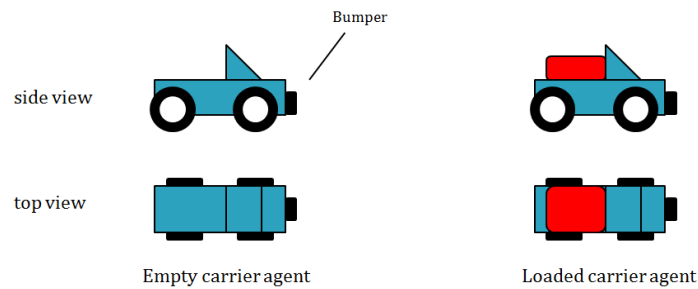

**Figure-1**: Carrier agent (side and top view)

2) *Loader / Un-loader agent:* It receives/delivers material from the storehouse, can detect if a carrier is waiting (for loading or unloading) by reading the presence sensor, it ensures that the carrier waiting to be loaded is loaded and the carrier waiting to be unloaded is unloaded.

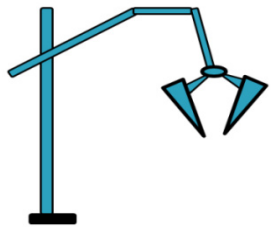

**Figure-2**: Loader / Un-loader agent

3) *Store-manager agent:* manages the stock material count in the storehouse and informs the loader agent or un-loader agent about the count.

The main purpose of KitRobot is to graphically program these three types of robotic agents.

## 2. Objectives

The principal objective is the development of integrated program development environment named KitRobot to create programs for these multi-agent robots. It involves a graphical interface for program development with the help of graphical symbols. A multi-agent robotic system involves a number of agents capable of mutual interaction, working together to carry out a complex task. In multi-agent robotic system a set of agents work together and carry out tasks, even if some agents of the multi-agent system fail than the work is distributed to other agents and they recovers the system whereas if a central system fails than it cannot be recovered. It supports XML language to load at run time different types of objects for program development. The use of XML technology makes this interface dynamic and flexible. The objects stored in these xml files are called KitRobot primitives. These primitives provide input to the system.

## 3. State of the art

### 3.1. HoRoCoL (Homogeneous Robotic Component Language)

HoRoCoL is a language for programming atoms or agents. It has formal explicit way to represent the actions and synchronizations between agents for the execution of a task. It is suited for reconfigurable modular robotics like a multi-agent robotic system. Programming such reconfigurable systems is a difficult task, as there are concepts like: methods or algorithms for planning and trajectory generation, or classical hierarchical architectures for robot control. There are different paradigms of robots: functional, deliberative or declarative, and synchronous. In any way, we can schematically summarize the difficulties of robot programming in two great characteristics:

a) Programming of elementary primitive actions of a robot is often a program including many process running in parallel with real-time constraints and local synchronization

b) Interactions with the environment are driven via traditional features, like interrupt on event or exception and synchronization with another element.

A multi-agent robotic system has a team of robot where cooperation and coordination are needed, and this introduces an additional difficulty of programming the behavior of a group of robots or even a society of robots. In this case programming implies to load a specific program on to each robot and each robot has different characteristics, sometimes different hardware, different behaviors and even different programming languages. These distinct programs must in general be synchronized to carry out missions of group and have reconfiguration capabilities according to a map of cooperation communication. From the human point of view, HoRoCoL is driven by three levels of team programming: Social, Group, and Agent. Social and Agent programming are very classical aspects. At the team level we focus on a specific group of robot. The original part of HoRoCoL work is on the Group programming where two new instructions SeqOfPar and ParOfSeq

a) **SeqOfPar**: The code is executed synchronously; it means that all agents execute their instruction at the same time. A control structure in which each line of the agent internal program will be executed synchronously over all concerned agents of the branch.

b) **ParOfSeq**: The code is executed independently in parallel; it means that each agent concerned by the code executes its code in parallel and there is no synchronization with any other agent. The only synchronization is at the end of ParOfSeq because it is terminated when all the agents concerned by the internal code have finished their execution [Duhaut et al., 2006] [Guyadec et al., 2005a] [Guyadec et al., 2005b].

## 3.2. Low Level Language

We have designed and developed an intermediate language with a well-defined grammar. This language is named Low Level Language. The application compiler translates the graphical program into this Low Level Language. This Low Level Language program is then transferred into the robot. There is also an interpreter designed and developed to interpret this Low Level Language code. This interpreter runs directly on our robot microprocessor.

**Table-1:** Low-Level Language Grammar

| Prog | Instructions |
|---|---|
| Instructions | instr ; { instr }* |
| Instr | action \| <br> action_interrupt \| <br> repeat \| <br> while \| <br> parallel \| <br> condition \| <br> event \| <br> timer \| <br> BREAK |
| Action | ident.ident ( param0/1 ) |
| action_interrupt | ° ident.ident ( param0/1 )° |
| integer_action | ident.ident ( param0/1 ) |
| param | variable { , variable }* |
| variable | nb \| integer_action \| condition |
| repeat | nb *( instructions ) |
| while | *[ cond ]( instructions ) |
| condition | [ cond ]( instructions ) <br> { !( instructions ) }0/1 |
| event | < cond >( instructions ) |
| cond | condition \| <br> !( cond ) \| <br> ( cond ) \| <br> cond & cond \| <br> cond \| cond |
| condition | ident.ident ( param0/1 ) |
| parallel | //( branches ) |
| branches | instructions , { instructions }* |
| timer | WAIT( nb ) |

Note:  ident    represents an identificator (i.e. a chain of characters)  
       nb       represents a whole number  
       *        is used for multiple times (zero or many)  
       0/1      is used for zero or one time

## 4. KitRobot: Requirement specification

During requirement specification the complete architecture of the system is analyzed. The goal is to select a solution that is best suited according to the needs and requirements. Each system has sub-systems, and for each sub-system solutions are chosen that are compatible with other fellow sub-system. The solutions that is better suited for integrating sub-systems into a complete application. First of all, to program any system a language is needed that has a well-defined grammar and syntax. So for the application KitRobot the first thing was to define a grammar for the generated code. Secondly, the program development primitives and robot object primitives are defined, and then the architecture specifications of KitRobot, functional as well as technical architecture are defined.

**4.1. Graph concept**

A program is developed in the form of a graph. A graph consists of vertices and arcs. Each graph contains at least two vertices (i.e. start and stop). Each vertex is connected to the other by an arc. Each vertex is a program element, which can be a constructor or a robot-object. These program elements are stored in the form of XML file. Each program element is loaded into the graphical interface at run time. XML format is chosen because of its open-standard, compatibility and universal portability. The program constructors which are the basic building blocks in developing a program are represented in XML file format.

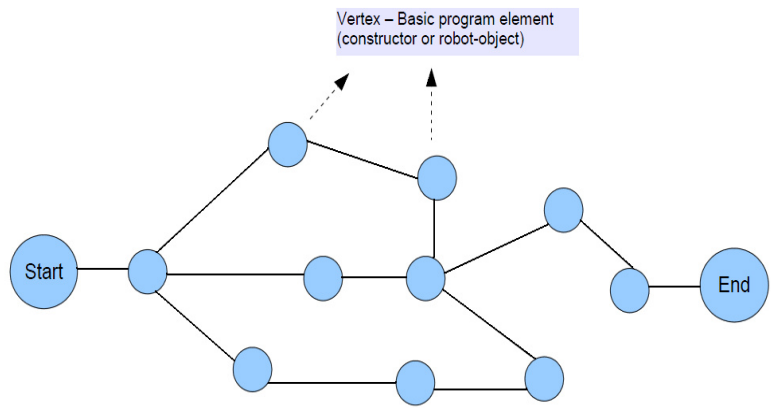

**Figure-3**: A Graph represents a program with each vertex as a program element

The number of vertices used in a program depends upon the complexity of the program.

**4.2. Program constructors**

Program constructors make the application flexible and allow further addition of features in the application without changing the source code. They are program building units, in other words are basic instructions for the development of a program. They are loaded at runtime into the KitRobot graphical interface from xml file. Many other constructors can be added depending upon the structure of robot and the programmer. They consists of Repetition, While, If Then, If Then Else, Wait, Break, And, Or, No, Parallelism, Branch, and Interrupt.
The Interrupt construct is used with parallelism. It can be added in a parallel flow of controls, and any branch having interrupt operation is forced to finish or abandon its task if all other branches have finished their respective tasks.

**4.3. Robot objects**

Robot objects are robot parts which are dependent on the hardware structure of the mini-robots. They are stored in the form of an xml file, and like program constructors they are also loaded at runtime into the graphical programming interface from this xml file.
Robot objects consist of actuators, sensors and variables.
1) **Actuators:**

*Actuator* is the mechanism by which an agent acts upon its environment. The agent can be either an intelligent software agent or any other autonomous being e.g. human, animal etc. Some examples of actuators of these various agents include: robot agent - grasping mechanism, electrical motor, hydraulic piston, wheel, door, light, switch etc. Examples of actuators defined in our application are:

| Actuator | Procedure |
|---|---|
| **Motor** | AbsoluteTurn(angle, speed) <br> Turn(speed) <br> Stop() <br> Where the angle and speed are arguments. The possible values of angle and speed must be between 0 and 100 |
| **Wheel** | Advance(speed) <br> Reverse(speed) <br> Stop() <br> Both Advance and Reverse have one argument speed. The value of speed must be between 0 and 100 |
| **Door** | Open() <br> Close() |
| **Light** | LightOn() <br> LightOff() |
| **Switch** | SwitchOn() <br> SwitchOff() |

2) **Sensors:**

A *sensor* object senses for any external stimulus. They are also referred to as conditions that are to be satisfied to carry out an action by the actuators e.g. captors, laser range finder, button etc. They are also referred to as conditions that are to be satisfied to carry out an action by the actuators. They include:

| Sensor | Operations |
|---|---|
| **Captor** | It is simple sensor and three operations are defined for it each operation with one parameter <br> EqualTo(value) <br> LessThan(value) <br> GreaterThan(value) |
| **Light** | It can also act like a sensor and we have defined three possible options for light sensors <br> IsRed() <br> IsGreen() <br> IsBlue() |
| **Button** | It has two options either the button can be pressed or released <br> IsPressed() <br> IsReleased() |

3) **Variables:**

A *variable* is an important part of program development. Each variable has a name and a value assigned to it. It is identified by its name. It can be used in different parts of a program with actuators and sensors e.g. integer variable, boolean variable etc.

They are important part of program development. Each variable has a name and a value assigned to it. Variable is identified by its name. They are used in different parts of a program with Actuaters and Sensors. I have defined two types of variables for the kitRobot programs. They are:

| **Integer variable** | It can be assigned a value between 0 and 100 |
|---|---|
| **Boolean variable** | It can be assigned *true* or *false* <br> Each robot has its specific Robotobject xml file whereas the constructors xml file remains the same for all robots. |

**5. KitRobot: Architecture**

The development architecture that is best suited to the needs and requirements of the user is defined. The functional as well as technical architecture is defined.

**5.1. Functional Architecture**

The functional architecture of the application is defined. This architecture consists of three smaller functional parts. Each part has a clearly defined function independent from other parts.

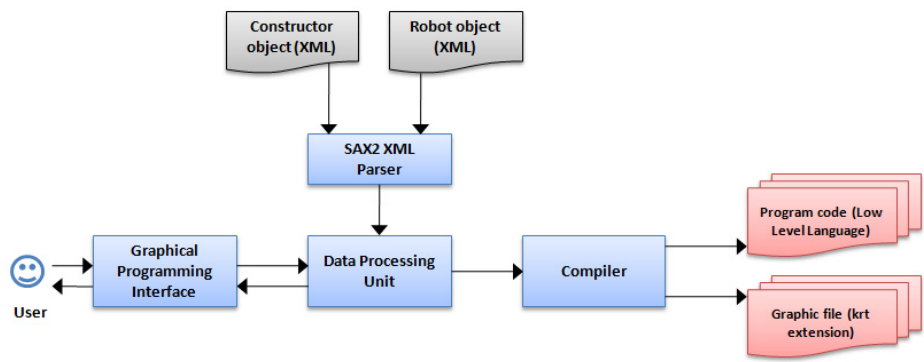

**Figure-4**: Functional architecture

Constructor object XML stores program building elements; Robot object XML stores robot parts, they are dependent on the hardware structure of the Robot; SAX2 XML Parser parses the constructor and robot object files and transfers the data stored in these xml files into the graphical interface. This XML data is then used for the program construction. Graphical Programming Interface manages a simple interface that allows programmer to program graphically by pick and drop; Data Processing Unit is the main control unit that manages all other units. It takes xml data from XML parsers and also user input from the Graphical Interface. Provides input to the data processing unit which generates Low Level Language code generator portion for the generation of Low Level Language code. This Low Level Language code is then transferred into each robot where it is converted into direct executable code with the help of a robot interpreter, and subsequently executed. This portion can also generate a KitRobot program file with *krt* file extension and stores the graphical program.

**5.2. Technical architecture**

The Technical Architecture is divided into three main parts:

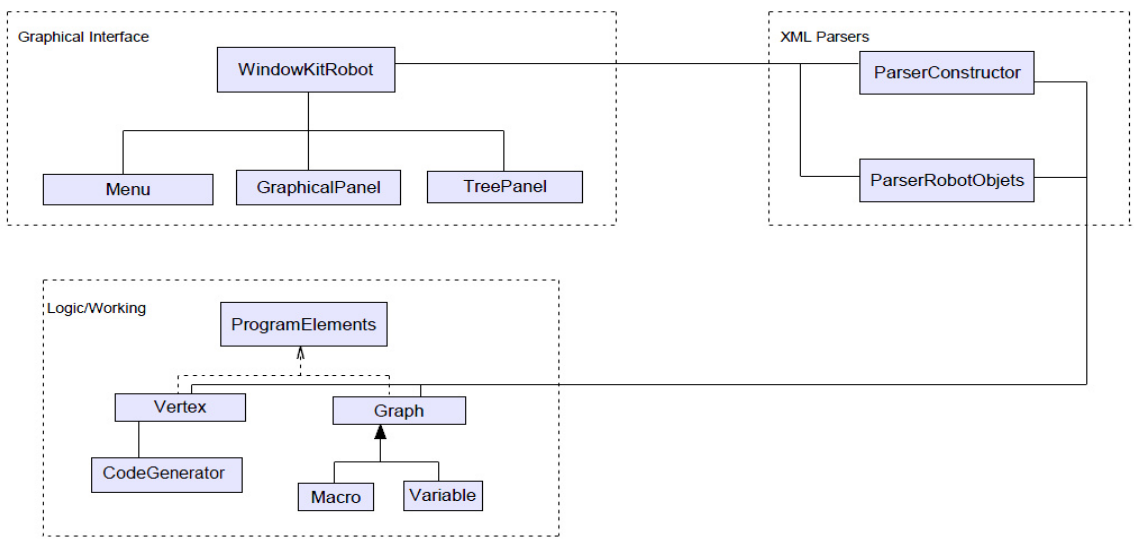

**Figure-5**: Technical architecture

1) Graphical Interface consists of classes that allow the user to program graphically by pick and drop. WindowKitRobot manages all other classes and connects them to the other component of the application. Menu manages all elements of menu bar and its eventHandling. TreePanel is the pick and drop panel on the right-side of KitRobot that contains the program constructors and robot objects. GraphicalPanel is the main drawing panel of the KitRobot.

2) XML Parsers are program constructors and robot objects are stored in the form of xml files. XML parsers are for these two types of xml files. Program constructors are program building units. Robot objects are robot parts and are totally dependent to the hardware structure of the Robot.

3) Processing Unit is the main working unit of the KitRobot. Program is constructed and stored in the form of a Graph. Each program is in fact a Graph containing multiple vertices. Each vertex is a program construct which can be a constructor or a robot object. CodeGenerator works as a compiler, it converts the graph into a low level code which can be executed on the robot.

## 6. Conclusion

Our major achievement is the definition and implementation of grammar for the Intermediate language. This grammar defines the Intermediate language. This Intermediate language has a three-address code format. The graphical program is translated into this three-address code based Intermediate language. The second goal attained is the definition of Integer and Boolean variables for graphical programming based KitRobot. The third milestone achieved is the definition and implementation of actuator and sensor constructs for KitRobot graphical programming.

**Acknowledgement**

We are grateful to Prof. Dr. Muhammad Mukhtar, Honourable Vice Chanceller, The Islamia University of Bahawalpur for his continuous encouragement and support for computer science research projects.
We are also thankful to Prof. Dominique Duhaut of IRISA - University of South Brittany, Bretagne, FRANCE.